\pdfoutput=1

\documentclass[10pt,twocolumn,letterpaper]{article}

\usepackage{cvpr}
\usepackage{tikz}\usetikzlibrary{positioning,arrows.meta,calc}

\usepackage{tikz}\usetikzlibrary{positioning,arrows.meta,calc}

\providecommand{\FloatBarrier}{}

\definecolor{cvprblue}{rgb}{0.21,0.49,0.74}
\usepackage[pagebackref,breaklinks,colorlinks,allcolors=cvprblue]{hyperref}
\hypersetup{pdftitle={Disentangling EEG Signals for Visual Decoding},pdfauthor={Marisa Ripoll, Moritz Nicolas Andres, Abbas Sohrabpour, Sheraz Khan, Klaus Diepold}}

\def\confName{CVPR}
\def\confYear{2027}

\title{Disentangling EEG Signals for Visual Decoding}

\author{
Marisa Ripoll$^{1,2}$ \quad
Moritz Nicolas Andres$^{1, 2}$ \quad
Abbas Sohrabpour$^{1}$ \quad
Sheraz Khan$^{1}$ \quad
Klaus Diepold$^{2}$\\[4pt]
$^{1}$Athinoula A. Martinos Center for Biomedical Imaging, Massachusetts General Hospital,\\
Harvard Medical School, Charlestown, MA, USA\\
$^{2}$Technical University of Munich, Germany\\
{\tt\small marisa.ripoll@tum.de}
}

\begin{document}
\maketitle
\begin{abstract}
In EEG-based visual decoding, when images are shown in quick succession, the brain is still processing one image when
the next arrives. Every second of EEG recorded this way therefore contains the brain responses to several images
at once. Despite this fact, EEG decoders treat each window as the response to a single image and treat the rest as
noise. We take the opposite view and build on the premise that the signal is an overlap of the brain's responses to
different stimuli. We derive a mathematical representation of the signal in which the input window corresponds to the
sum of the responses to all images being processed. Each response is shifted by the moment of appearance and
transformed spatially as it propagates through the brain. We turn this representation directly into a decoder.
The resulting decoder distinguishes several images from a single window, including images that published decoders cannot
detect at all. On single trials, as in live BCI decoding, it is more accurate than every published model we evaluated, in
all ten subjects (16.8\% vs.\ 14.1\% top-1 for the strongest on THINGS-EEG2), and it reaches 68.3\% top-1 on the standard benchmark and 71.5\% when decoding each repetition before averaging, with only 4.1\,M parameters. Because each part of the decoder corresponds to a term of the signal representation, the model becomes interpretable.
\end{abstract}
    
\section{Introduction}\label{sec:intro}
Publications applying AI to neuroscience have grown roughly fivefold in the past
decade~\cite{tekinBibliometricAnalysisStudies2025,zhouBirdsEyeView2025} and faster still for deep learning for
neuroimaging~\cite{chakiDeepLearningRevolution2026}. EEG is a striking example. It is noisy and spatially blurry, and
was long considered too crude for fine-grained visual decoding, yet new models now appear every few months, each
beating the last on the same benchmarks. On THINGS-EEG2~\cite{giffordLargeRichEEG2022}, top-1 accuracy rose from
5.82\%~\cite{duDecodingVisualNeural2023} to 28.5\%~\cite{liVisualDecodingReconstruction2024} in a single year, then to
50.9\%~\cite{wuBridgingVisionBrainGap2025} and, within the last year, to
63.2\%~\cite{zhangNeuroBridgeBioInspiredSelfSupervised2026} and 80.0\%~\cite{liuLeveragingVisualBlur2026}, with preprints
reporting higher numbers still~\cite{liuFSDBNForegroundAwareEEGVisual2026,wuComplementarityDrivenDualTeacherHierarchical2026}.
These gains come with increasingly complex architectures that act as black boxes.

\begin{figure*}[t]\centering
\includegraphics[width=0.92\linewidth]{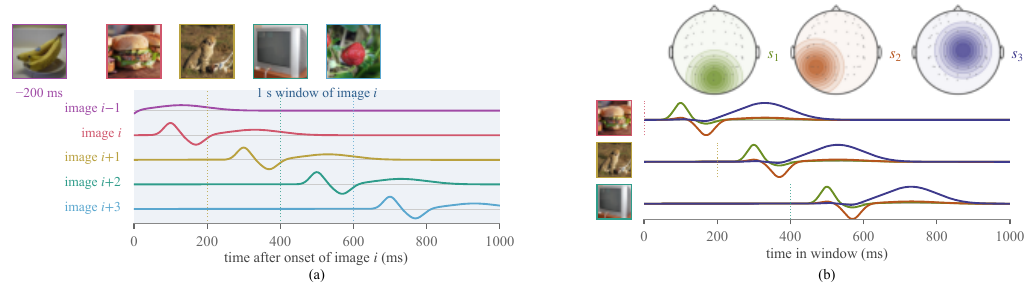}
\caption{\textbf{Several images in one window} (schematic). (a) Images appear every $d = 200$\,ms, so the one-second window of image
$i$ contains the end of the response to image $i{-}1$ and the beginning of the responses to later images. (b) The
recording is a sum of sources, each with a fixed topographical map (top). Every image evokes its own response in each
source, shifted by the moment it appeared (bottom).}\label{fig:window}\end{figure*}

EEG recordings contain a large amount of noise. To reduce it, datasets usually show each image many times and average
the recorded brain responses per subject, which removes activity unrelated to the image. THINGS-EEG2 shows each test
image 80 times, and the standard benchmark decodes this 80-repetition average. However, a brain-computer interface
working in real time has no such luxury. It must decode what a person sees from a single presentation.
We refer to this as the single-trial setting. Accuracies here are far lower, because all the noise is left in, but real applications depend on this setting.

The need for many repetitions also shapes how the data is recorded. To collect as many trials as possible, recent datasets use rapid serial visual presentation (RSVP)~\cite{giffordLargeRichEEG2022,grootswagersRepresentationalDynamicsVisual2019},
showing images in quick succession. The brain, however, takes far longer to process an image than RSVP allows. In
THINGS-EEG2 a new image appears every 200\,ms, while decoding models analyze the full second after each onset. By the
end of that second the subject has seen four more images, and the brain is still processing the previous one. This raises our first motivating question.
\begin{center}\emph{How many images does the brain hold at once?}\end{center}
To answer this question we conducted a pilot study (Supp.~\ref{sec:supp_leak}), where we trained a linear classifier on
these windows. The classifier was able to identify the current image among 200 categories with 10.7\% accuracy, and it
also identified the previous image (5.8\%) and the next three (8.7\%, 6.2\% and 4.1\%). Every window therefore carries
the identity of five images, not one.

If several images can be distinguished within the same window, a second question follows.
\begin{center}\emph{How can the images in one window be told apart?}\end{center}
Visual processing unfolds in stages, from early visual cortex, which responds to edges and contrast, to higher regions
that encode objects and meaning~\cite{fellemanDistributedHierarchicalProcessing1991}, a hierarchy mirrored by the layers of deep neural networks~\cite{cichyComparisonDeepNeural2016}. An image that appeared 600\,ms ago and one that has just
appeared are therefore at different stages of the same process (Fig.~\ref{fig:window}b). Current decoding models
ignore this, treating each window as the response to one image and everything else as noise.

\begin{figure*}[t]\centering
\includegraphics[width=0.9\linewidth]{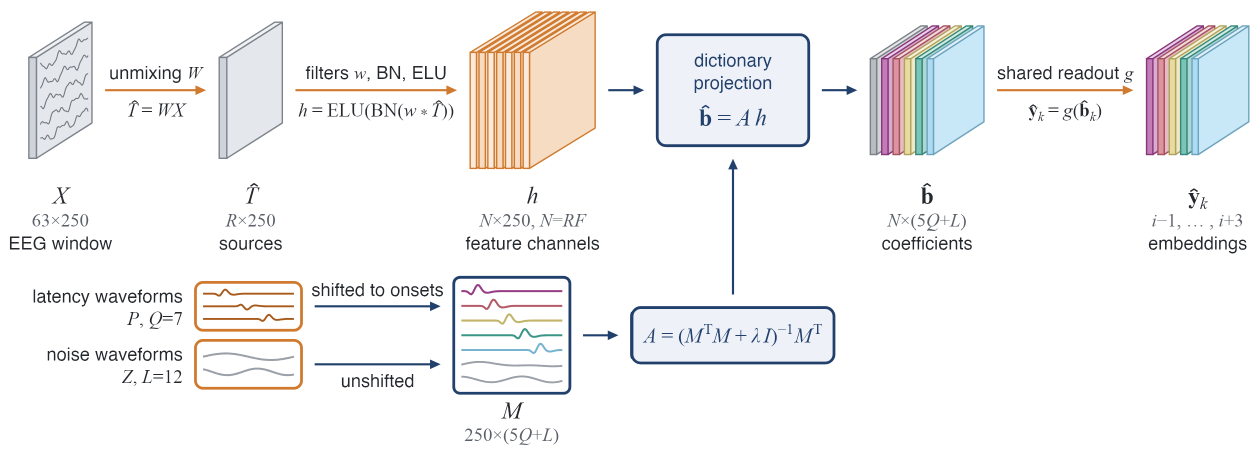}
\caption{\textbf{Architecture.} The EEG window is unmixed into sources and filtered into feature channels $h$. The
dictionary $M$ places the learned latency waveforms $P$ at the onsets of the five images and adds the noise waveforms
$Z$. The dictionary projection returns each image's coefficients $\hat{\mathbf b}_k$, and one shared readout maps them
to an embedding per image. Noise coefficients (gray) are not read out. Orange marks learned parameters, navy closed-form
computation, and colors images $i{-}1,\dots,i{+}3$. Sizes are for one window.}\label{fig:model}\end{figure*}

We take the opposite view and build our architecture directly from this model of the brain. Every window is modeled as the
sum of the responses to all the images being processed, each shifted by the moment it appeared and expressed through a
small set of learned neural sources (Fig.~\ref{fig:model}). We fit this superposition to the EEG by least squares, using a dictionary of response waveforms that is learned from the data and shared across participants, and map each image's contribution to an image embedding with one linear layer. What does this
architecture bring?
\begin{itemize}
\item The model is capable of decoding all five images in a window. Even when trained with the label of the current
image only, it retrieves the other four at 9.8\% top-1 on average, where published decoders are at chance
(Sec.~\ref{sec:multi}).
\item On single trials, it reaches 16.8\% top-1, above the best published model we evaluated (14.1\%) in
all ten subjects, and 68.3\% on the 80-repetition benchmark (Sec.~\ref{sec:perf}).
\item Finally, every component of the model corresponds directly to a term in our representation of the brain signal,
so the model can be easily interpreted (Sec.~\ref{sec:interp}).
\end{itemize}

\section{Related work}
Contrastive alignment of EEG with pretrained image embeddings has become the standard approach to visual decoding on THINGS-EEG2~\cite{giffordLargeRichEEG2022}. NICE~\cite{songDecodingNaturalImages2024a} pairs a temporal and spatial convolution with CLIP. ATM~\cite{liVisualDecodingReconstruction2024} adds channel attention and reconstructs images with a diffusion model. UBP~\cite{wuBridgingVisionBrainGap2025} blurs the image targets according to an uncertainty estimate, and NeuroBridge~\cite{zhangNeuroBridgeBioInspiredSelfSupervised2026} adds augmentation priors on both sides. AVDE~\cite{daiAutoregressiveVisualDecoding2026} fine-tunes the LaBraM foundation model~\cite{jiangLargeBrainModel2024}. LVBP~\cite{liuLeveragingVisualBlur2026} aligns the EEG with features of each image at several levels of blur, following the blurred perception of fast presentation, and HVF~\cite{zhengLearningBrainRepresentation2026} aligns it with hierarchical features from several image encoders. FSDBN~\cite{liuFSDBNForegroundAwareEEGVisual2026} models dynamic brain networks. All of them map one window to one embedding of one image, are trained on 4-repetition averages and are evaluated on 80-repetition test averages. Their gains have come from larger encoders, better image targets and priors on the image side. None model the sequence of images that produced the window, and single-trial decoding, the setting of brain--computer interfaces, is rarely reported.

It is well established that fast presentation leaves several images in the brain at once. Decoding studies of RSVP
showed that each image can be decoded for about a second after its onset, so that successive images are represented
simultaneously~\cite{grootswagersRepresentationalDynamicsVisual2019}, and that the brain multiplexes them along the
visual hierarchy as traveling waves~\cite{kingHumanBrainEncodes2021}. These studies use the overlap to study
perception. We build on the same observation and use it to decode every image in a window with a model
of the overlap.

The overlap of successive responses is an old problem in ERP research. The ADJAR
technique~\cite{woldorffDistortionERPAverages1993} estimates and subtracts the overlap of adjacent responses from averages. Regression-based deconvolution (rERP~\cite{smithRegressionbasedEstimationERP2015},
Unfold~\cite{ehingerUnfoldIntegratedToolbox2019}) fits all overlapping responses jointly and can include item-level and
continuous predictors. RIDE~\cite{ouyangResidueIterationDecomposition2011} separates components by their latency
variability across trials, and temporal response functions relate neural signals to continuous stimuli by regularized
regression~\cite{crosseMultivariateTemporalResponse2016}. These methods are built to estimate response waveforms. The overlap correction used in decoding pipelines subtracts a response shared by all images, which removes the overlap from
the average but not the image-specific information a decoder needs (Supp.~\ref{sec:supp_leak}). Our model keeps the
regression view where a window is a sum of responses shifted by known onsets, solved jointly,
but gives every image its
own response, built from shared latency waveforms, learns the feature space in which the sum holds, and trains the whole
system for retrieval through the dictionary projection. Differentiable closed-form and optimization layers have been used for few-shot
learning~\cite{bertinettoMetalearningDifferentiableClosedform2018} and structured
prediction~\cite{amosOptNetDifferentiableOptimization2017}. Ours is a ridge regression whose design matrix is the model's own learned dictionary of waveforms (Sec.~\ref{sec:method}). Like hyperalignment~\cite{haxbyCommonHighDimensionalModel2011} and
the shared response model~\cite{chenReducedDimensionFMRIShared2015}, the model maps subjects into a common space, and only the
topographical maps are subject-specific.
 
\section{Method}\label{sec:method}
We build our model on five assumptions about the generation of EEG signals by spatiotemporal brain processes. These assumptions progressively
define our proposed decoder, which consists of two parallel branches that are brought together to separate overlapping image responses. The first one transforms the recorded EEG window into feature channels representing an approximation of the brain activity elicited by a continuous input of visual stimuli. The second one builds a parameterized dictionary of response waveforms that model how this activity can be explained by individual image responses. The two branches are then combined by fitting the response waveforms to the feature channels of the recorded signal with least squares. In this manner, we explain the spatiotemporal generation and propagation of brain activity elicited by each image with a shared dictionary learned from the data and features learned for each participant. This separates the contribution of each image, whose coefficients we finally map to image embeddings.

\paragraph{Assumption 1.} \textbf{The EEG signal X is the superposition of a limited number of underlying brain sources.}
The response to an image propagates through the brain, from early visual areas to higher-order ones, so that at any given moment a network of brain regions is processing the incoming visual stimuli, according to the sequence of the images and the hierarchy of the visual system. The electrodes on the scalp pick up all of these regions at once. The recording can therefore be described as a sum of sources, each with its own signal, reaching the electrodes with a fixed and unique spatial fingerprint in the scalp EEG recordings, i.e., its topographical map (topo map) (Fig.~\ref{fig:window}b), plus noise,
\begin{equation}
X=\sum_{r=1}^{R} s_r\,T_r+\eta=S\,T+\eta ,
\label{eq:sources}\end{equation}
where $s_r$ is the topographical map of source $r$, $T_r$ its signal, and $\eta$ is noise. To get the source signals back we aim at canceling the mixing. $S$ is unknown and not square, so we learn a matrix $W$ that takes the role of its pseudo-inverse $S^{+}$,
\begin{equation}
\hat T=S^{+}X\approx W X ,
\label{eq:unmix}\end{equation}
where $\hat T$ is the approximation of $T$. Head geometry differs between participants, so the unmixing matrix $W$ is estimated for each participant, while every other element of the model is shared among all of them.

The raw responses contain oscillatory bursts whose phase varies between trials, so that their positive and negative half-waves cancel when compared across trials. Letting the topographical maps absorb the sign of the EEG signals, we model only the presence and amplitude of this activity. We therefore filter each estimated source with a short learned filter $w$, normalize it, and rectify it with an ELU, which turns each burst into a positive amount of activity with a consistent timing, resulting in following equation for the feature channels of the first branch,
\begin{equation}
h(t)=\mathrm{ELU}\big(\mathrm{BN}(w\ast\hat T_r)\big)(t),
\label{eq:features}\end{equation}
where $\mathrm{BN}$ is batch normalization and $h(t)\in\mathbb R^N$ holds $N=R\times F$ channels, $F$ filters per source.

\paragraph{Assumption 2.} \textbf{Each source signal $T(t)$ is the sum of its responses to all images in view.}
In fast RSVP a new image appears every $d=200$\,ms, long before the response to the previous one has ended. Each source
thus responds to several images at once, each response starting when its image appeared (Fig.~\ref{fig:window}a),
\begin{equation}
T(t)=\sum_{k=-1}^{3} r_k(t-kd)+\eta,
\label{eq:superpos}\end{equation}
where $r_k$ is the response to the image at sequence position $k$, that is, its place in the presentation order relative to the current image, with $k=0$ the current image, $k=-1$ the previous one and
$k=1,2,3$ the next three.

\paragraph{Assumption 3.} \textbf{Assumption 2 still holds after filtering and rectification.}
Filtering keeps the sum in Eq.~\eqref{eq:superpos}, and we assume the ELU in Eq.~\eqref{eq:features} roughly does too.
Each channel is then a sum of the responses of the images, each shifted to its onset, plus noise,
\begin{equation}
h(t)\approx\sum_{k=-1}^{3} f_k(t-kd)+\eta,
\label{eq:featsum}\end{equation}
where $h(t)\in\mathbb R^N$ and $f_k$ is the response of the channel to the image at position $k$, that is, how its
activity rises and falls after the image appears.

\begin{figure}[t]\centering
\begin{minipage}[b]{0.31\linewidth}\centering\includegraphics[width=0.837\linewidth]{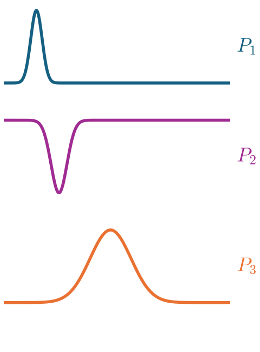}\\(a)\end{minipage}\hfill
\begin{minipage}[b]{0.67\linewidth}\centering\includegraphics[width=0.827\linewidth]{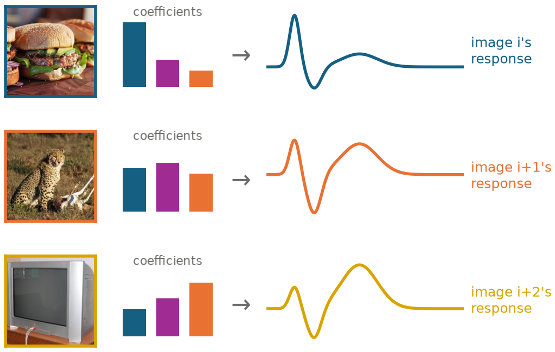}\\(b)\end{minipage}
\caption{\textbf{Responses from shared waveforms} (schematic). (a) Latency waveforms $P_q$ shared by all images,
here three illustrative ones (the model learns $Q{=}7$). (b) Each image mixes them with its own coefficients $\mathbf b_k$, giving its response
$f_k$ (Eq.~\ref{eq:response}).}\label{fig:concept}\end{figure}

\paragraph{Assumption 4.} \textbf{All responses $f_k$ are mixes of a few fixed waveforms.}
All images pass through the same visual system, so their responses rise and fall at similar latencies and differ mainly in their contribution to each channel. A free curve per image and channel would also have too many unknowns for one window. Instead, every response mixes the same $Q$ latency waveforms $P_q$ with coefficients $b_{k,q}$ that depend on the image (Fig.~\ref{fig:concept}),
\begin{equation}
f_k=\sum_{q=1}^{Q} b_{k,q}\,P_q=P\,\mathbf b_k ,
\label{eq:response}\end{equation}
where $P$ holds the waveforms as columns and $\mathbf b_k$ the coefficients of image $k$. The waveforms are shared, and the coefficients
tell the images apart.

\paragraph{Assumption 5.} \textbf{The noise is a mix of a few fixed waveforms.}
Each channel also contains activity unrelated to the images, such as ongoing rhythms and slow drifts. This noise can be understood as brain processes irrelevant to the visual processing of the images, happening in the background. It is not tied to
any onset, so we describe it with $L$ unshifted noise waveforms $Z_l$ and coefficients $a_l$.
\begin{equation}
\eta\approx\sum_{l=1}^{L}a_l\,Z_l(t),
\label{eq:noise}\end{equation}

\paragraph{The model} Combining Assumptions 4 and 5 with Assumption 3, i.e., substituting Eqs.~\eqref{eq:response}
and~\eqref{eq:noise} into Eq.~\eqref{eq:featsum}, we get the per-channel final representation of signal $h(t)$, i.e.,
the dictionary of response waveforms of the second branch,
\begin{equation}
\begin{aligned}
h(t)&\approx\sum_{k=-1}^{3}\sum_{q=1}^{Q} b_{k,q}\,P_q(t-kd)+\sum_{l=1}^{L}a_l\,Z_l(t)\\
&=\sum_{k=-1}^{3} P_k\,\mathbf b_k+Z\mathbf a=M\mathbf b ,
\end{aligned}
\label{eq:model}\end{equation}
where $P_k$ is $P$ shifted to the onset of image $k$ and cut at the window edges. The dictionary $M=[P_{-1},\dots,P_3,Z]$
 has $5Q+L$ columns and is the same for every channel, and $b=[b_{-1};\dots;b_3;a]$
holds the coefficients of one channel.

Combining the two branches, we can substitute $h(t)$ in Eq.~\eqref{eq:model} with Eq.~\eqref{eq:features}. Since $M$ is
computed from the waveforms, this leaves the coefficients in Eq.~\eqref{eq:model} as the only unknowns. The equation is linear in them, so we solve this inverse problem by least squares, regressing the image contribution coefficients by projecting the feature channels $h$ onto the waveform dictionary $M$. We call this step the dictionary projection,
\begin{equation}
\begin{aligned}
\hat{b}&=A\,h\approx M^{+}h,\\
A&=(M^\top M+\lambda I)^{-1}M^\top.
\end{aligned}
\label{eq:solve}\end{equation}
The waveforms of neighboring images overlap in time, so the same activity could be attributed to more than one image. The inverse $(M^\top M + \lambda I)^{-1}$ accounts for this overlap, so that each part of a channel is assigned once, to the image or noise waveform that best explains it. The small $\lambda$ is a regularization constant necessary for the stability of the pseudo-inverse. $A$ has no parameters of its own and acts as a differentiable closed-form
layer~\cite{amosOptNetDifferentiableOptimization2017,bertinettoMetalearningDifferentiableClosedform2018}.

We decode each image from its channel coefficients. The noise coefficients are dropped since they are only needed for
the least squares solver, the coefficients of image $k$ from all $N$ channels are stacked into a vector
$\hat{\mathbf b}_k$ of length $NQ$, and one linear readout $g$ maps it to an image embedding,
\begin{equation}
\hat{\mathbf y}_k=g(\hat{\mathbf b}_k),\qquad k=-1,\dots,3 .
\label{eq:readout}\end{equation}
$g$ is shared by all positions, so an image is processed the same way wherever it falls in the window.

\paragraph{Training.}
In a batch of $B$ windows, let $C_k$ be the $B\times B$ matrix of cosine similarities between the predicted embeddings
$\hat{\mathbf y}_k$ and the true embeddings $\mathbf y_k$ of the images at position $k$. A contrastive loss asks each
window to pick out its own image, and each image its own window,
\begin{equation}
\mathcal L=\sum_{k}\tfrac12\big[\mathrm{CE}(sC_k,I)+\mathrm{CE}(sC_k^\top,I)\big],
\label{eq:loss}\end{equation}
where CE is the cross-entropy, $s$ a learned temperature, and the sum runs over the current image and, when their labels
are used, the four neighbors. The waveforms $P_q$ and $Z_l$ are learned like every other parameter, from a random initialization and on all participants together. The contributions $\hat b$ are not learned. They are computed for each window by the dictionary projection. Training therefore adjusts the waveforms so that this projection keeps the images apart. Neighbor labels are used only here. At test time, one forward pass gives all five embeddings. Only 12.1 k of the 4.1 M parameters describe the EEG signal ($W$, the filters and the waveforms). The rest form the linear readout $g$ which maps the separated contributions to the image embedding.

We also tested whether a learned layer can replace the dictionary projection. With the labels of all five images, a learned matrix in place of $A$, initialized from the projection, reaches the same accuracy (Supp.~\ref{sec:supp_ablations}). But when only the label of the current image is available, a model without the projection cannot decode the neighbors at all, while ours can (Sec.~\ref{sec:multi}). The projection is therefore what separates the images, by assigning each response to its known onset.

\section{Experimental setup}\label{sec:setup}
We use THINGS-EEG2~\cite{giffordLargeRichEEG2022}, with ten subjects, 63 electrodes, 16,540 training images from 1,654 concepts seen four
times each, and 200 test images from 200 concepts not seen in training, shown 80 times each. Images were shown in fast
RSVP sequences, one every 200\,ms. We take windows of 0--1000\,ms after each onset at 250\,Hz, baseline-corrected, with
multivariate noise normalization fitted on training data only. For every window the image at each position is known,
so the neighbors' labels come for free.

The task is 200-way zero-shot retrieval, in which an EEG embedding is compared with the embeddings of all 200 test images, and we
report top-1 and top-5 accuracy (chance 0.5\% and 2.5\%). The classical protocol averages the 80 repetitions of each test
image before decoding, and the single-trial protocol decodes each of the 16,000 test trials on its own, as live decoding
would. The neighbors can only be scored on single trials, since each repetition of an image had different neighbors.
Every setting (architecture, epochs, targets) is chosen on 10\% of the training concepts, held out entirely
so that validation is zero-shot like the test, and the test set is scored once after the last epoch
(Supp.~\ref{sec:supp_settings}).

Our model has $R{=}32$ sources, $F{=}12$ filters of 13 samples (52\,ms), $Q{=}7$ waveforms and $L{=}12$ noise waveforms. Its
targets are the DINOv2-giant~\cite{oquabDINOv2LearningRobust2024} hidden state at layer 20 of 40, averaged over tokens. It is trained on all
subjects together for 12 epochs and then fine-tuned per subject for 4, with half of every batch replaced by 4-repetition averages of the same image and subject (Supp.~\ref{sec:supp_aug}). We call this training data the mix. One model, with three seeds, is used for
every result unless stated otherwise.

We compare against three of the strongest published models, UBP~\cite{wuBridgingVisionBrainGap2025},
NeuroBridge~\cite{zhangNeuroBridgeBioInspiredSelfSupervised2026} and LVBP~\cite{liuLeveragingVisualBlur2026}. The
authors' full models are retrained with their code and settings, and our reproductions land within 1.1 points of the
reported numbers. To compare architectures rather than training recipes, we also train the encoders of UBP and
NeuroBridge inside our
pipeline with our recipe (same data, targets, splits and pre-training, with their own losses and optimizers), both on
4-repetition averages, as published, and on single trials. The neighbors get one output head per position,
trained with the neighbors' labels like ours.

\section{Results}\label{sec:results}
All results are 200-way zero-shot retrieval on THINGS-EEG2, averaged over the ten subjects, with the test set scored
once. Ours are averaged over three seeds per subject (five for the model trained on the mix with the label of
image $i$ only), and $\pm$ is the standard deviation over subjects. Statistical tests are two-sided paired $t$-tests
over subjects.

\subsection{Decoding several images per window}\label{sec:multi}
\begin{table}[t]\centering\small\setlength{\tabcolsep}{3pt}
\resizebox{\linewidth}{!}{%
\begin{tabular}{lccccc}\toprule
Method & $i{-}1$ & $i$ & $i{+}1$ & $i{+}2$ & $i{+}3$\\\midrule
UBP & 0.3$\pm$0.0 & 5.0$\pm$1.2 & 0.3$\pm$0.0 & 0.5$\pm$0.1 & 0.5$\pm$0.0\\
NeuroBridge & 0.3$\pm$0.1 & 6.0$\pm$1.6 & 0.3$\pm$0.1 & 0.4$\pm$0.0 & 0.5$\pm$0.0\\
LVBP & 0.2$\pm$0.0 & 14.1$\pm$3.2 & 0.2$\pm$0.0 & 0.4$\pm$0.1 & 0.4$\pm$0.1\\\midrule
Ours & \textbf{2.4}$\pm$0.6 & \textbf{17.9}$\pm$4.1 & \textbf{13.9}$\pm$2.8 & \textbf{12.9}$\pm$2.7 & \textbf{10.1}$\pm$2.1\\\bottomrule
\end{tabular}}
\caption{\textbf{Neighbors without their labels.} Single-trial top-1 accuracy (\%) for each image in the window when
every model is trained with the label of image $i$ only. Published methods use the authors' code, and our model is
trained on single trials. Chance is 0.5\%. Our model is above chance at every position in all ten subjects
($p<10^{-5}$, one-sample $t$-test).}\label{tab:free}\end{table}

\begin{figure}[t]\centering\includegraphics[width=0.8\linewidth]{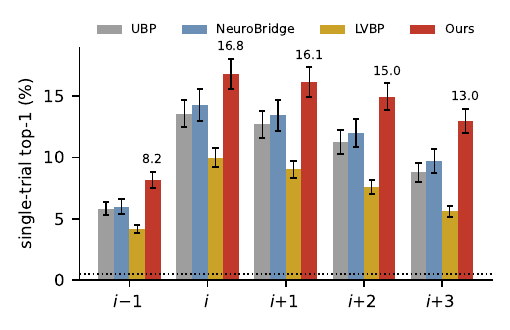}
\caption{\textbf{Five images from one window.} Single-trial top-1 accuracy for each image when every model is trained
with the labels of all five images. The baselines are their EEG networks with one output head per position, trained in
our pipeline. Parameters for all five images are 20.7\,M for UBP, 12.8\,M for NeuroBridge, 19.6\,M for LVBP and 4.1\,M for
ours. LVBP's network is trained per subject, as by its authors (three seeds). Dotted
line, chance. Ours is the most accurate at every position in all ten subjects ($p<10^{-4}$).}\label{fig:neighbours}\end{figure}

Our model computes coefficients for every image in the window, not just image $i$, so the readout can be applied to
all five of them. We first ask whether the neighbors can be decoded at all when no model has been given their labels. In
Table~\ref{tab:free}, every model is trained with the label of image $i$ only, as in all published work. The published
methods stay at chance on every neighbor. To predict image $i$, they learn to ignore everything else in the window.
Our model has never seen a neighbor's label, yet it retrieves the next three images at 13.9\%, 12.9\% and 10.1\% top-1
and the previous image at 2.4\%, whose early response falls before the window. The neighbors are decoded with the
readout learned for image $i$ alone, because the dictionary projection, the least-squares fit of every image's waveforms at its onset (Sec.~\ref{sec:method}), places each image's response in its own coefficients, in the
same form as those of the current image.

Removing the dictionary projection shows that it is responsible for this. For our model trained on the mix, applying the readout to the
features directly, with the same number of inputs, drops the next image from 12.3\% to 0.3\% and image $i$ from 16.2\%
to 9.6\%. The gain for image $i$ is not explained by the neighbors being modeled, because modeling fewer or more
images barely changes it (Supp.~\ref{sec:supp_ablations}). It comes from the latency waveforms, which describe each
response with a few numbers instead of a full time course.

One could argue that the published models were never asked to decode the neighbors. We therefore give their EEG
networks one output head per position, each trained with the labels of its image (Fig.~\ref{fig:neighbours}). Our model
keeps its single readout and is still the most accurate at every position, by 2.2 to 3.3 points, in all ten subjects. The margin is largest at $i{+}3$, where only the first 400\,ms of the response fall
in the window. Five separate NeuroBridge networks, one per position (48.4\,M parameters), also remain below it
(Supp.~\ref{sec:supp_recipe}). Learning the neighbors costs image $i$ 1.2 points on single trials (16.7\% against 17.9\%) for four more
images per window.

\subsection{Single-trial and averaged retrieval}\label{sec:perf}

\begin{table}[t]\centering\small\setlength{\tabcolsep}{4pt}
\begin{tabular}{llcc}\toprule
Method & Params & Top-1 & Top-5\\\midrule
UBP~\cite{wuBridgingVisionBrainGap2025} & 8.9M & 5.0$\pm$1.2 & 16.5$\pm$3.0\\
NeuroBridge~\cite{zhangNeuroBridgeBioInspiredSelfSupervised2026} & 9.7M & 6.0$\pm$1.6 & 18.9$\pm$4.0\\
LVBP~\cite{liuLeveragingVisualBlur2026} & 5.2M & 14.1$\pm$3.2 & 36.6$\pm$5.9\\\midrule
Ours, label of image $i$ & 4.1M & 16.2$\pm$3.9 & 40.1$\pm$6.7\\
Ours, labels of all five & 4.1M & \textbf{16.8}$\pm$3.9 & \textbf{41.0}$\pm$6.6\\\bottomrule
\end{tabular}
\caption{\textbf{Single trials.} Retrieval accuracy (\%) when every test trial is decoded on its own. Published methods
are run with the authors' code and settings. Our model is trained on the mix. Params, EEG side at
inference. Ours is more accurate than each published method in all ten subjects ($p<10^{-4}$ against LVBP, the
strongest).}\label{tab:single}\end{table}

\begin{table}[t]\centering\small\setlength{\tabcolsep}{4pt}
\begin{tabular}{lccc}\toprule
EEG network & Averages & Single trials & Mix\\\midrule
UBP & 7.6$\pm$2.2 & 11.8$\pm$3.4 & 10.1$\pm$3.1\\
NeuroBridge & 8.8$\pm$2.6 & 15.3$\pm$4.8 & 12.3$\pm$4.0\\
LVBP & 5.3$\pm$1.3 & 13.0$\pm$3.1 & 10.6$\pm$2.7\\
Ours & \textbf{12.8}$\pm$2.8 & \textbf{17.9}$\pm$4.1 & \textbf{16.2}$\pm$3.9\\\bottomrule
\end{tabular}
\caption{\textbf{Architectures under equal training.} Single-trial top-1 accuracy (\%) of the EEG networks of the
published methods and of our model, all trained in our pipeline with the same data, targets and label of image $i$, on
4-repetition averages, single trials or the mix (Supp.~\ref{sec:supp_recipe}). In every regime ours is the most
accurate in all ten subjects ($p<2\cdot10^{-4}$ against each network).}\label{tab:arch}\end{table}

\begin{figure*}[t]\centering\makeatletter
\begin{minipage}[t]{0.555\linewidth}\vspace{0pt}\centering\def\@captype{figure}
\includegraphics[width=\linewidth]{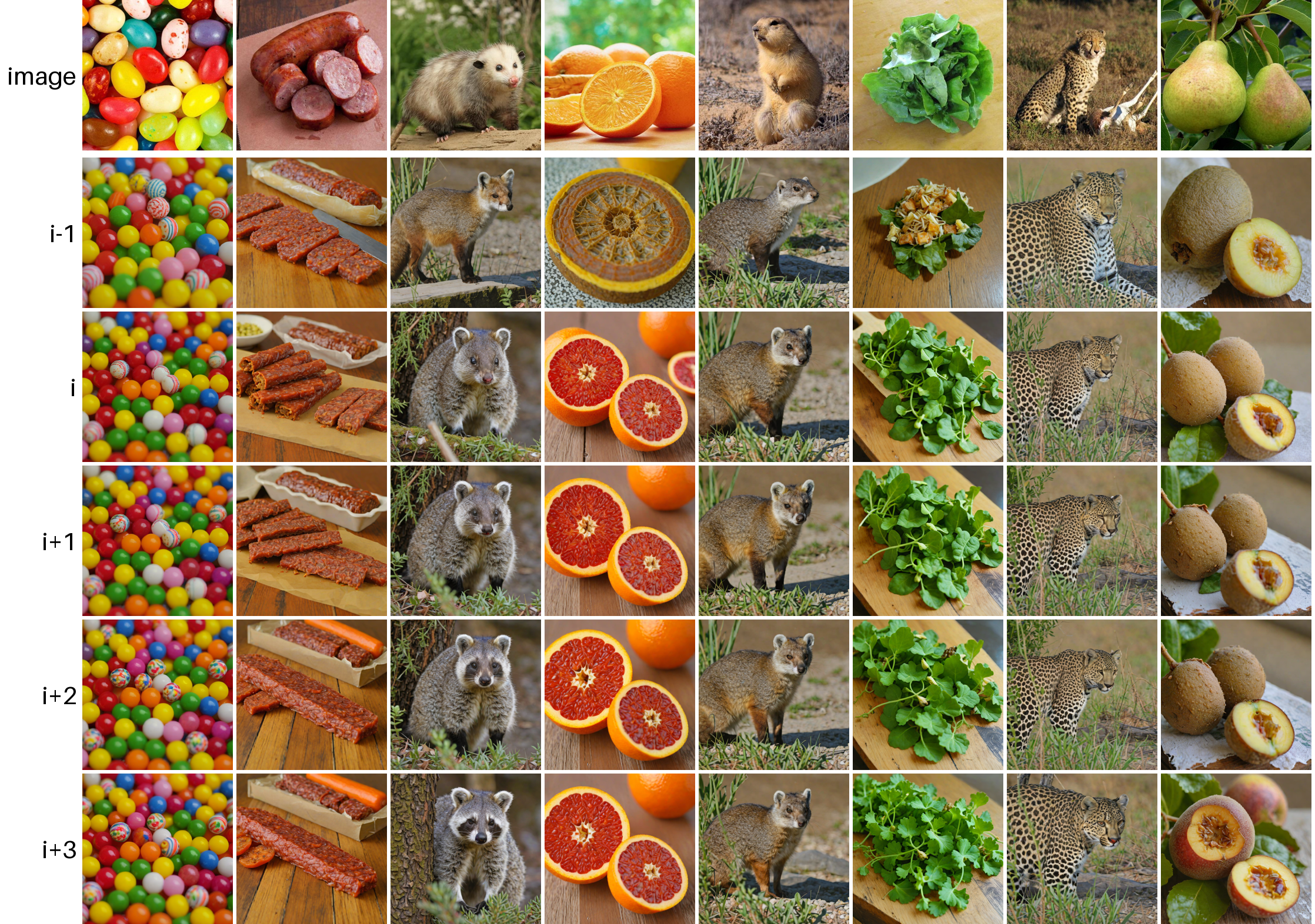}
\caption{\textbf{One image, read from every position.} Each column shows a test image (top) and its reconstruction from
the windows in which it appeared at position $i{-}1$, $i$, $i{+}1$, $i{+}2$ or $i{+}3$, averaging its presentations at
that position. The columns are the eight test images whose reconstructions are most similar to the image over the five
positions, out of 2,000 pairs of subject and image. Random examples and single presentations are in
Supp.~\ref{sec:supp_recon}. Reconstructions use SDXL-Turbo and IP-Adapter, as in
ATM~\cite{liVisualDecodingReconstruction2024}.}\label{fig:posavg}
\end{minipage}\hfill
\begin{minipage}[t]{0.42\linewidth}\vspace{0pt}\centering\def\@captype{table}\small\setlength{\tabcolsep}{4pt}\renewcommand{\arraystretch}{1.2}
\begin{tabular}{lcc}\toprule
Method & Top-1 & Top-5\\\midrule
NICE~\cite{songDecodingNaturalImages2024a} & 16.1 & 43.6\\
ATM-S~\cite{liVisualDecodingReconstruction2024} & 28.5 & 60.4\\
LaBraM~\cite{jiangLargeBrainModel2024}, as in AVDE~\cite{daiAutoregressiveVisualDecoding2026} & 30.0 & 58.2\\
UBP~\cite{wuBridgingVisionBrainGap2025} & 50.9 & 79.7\\
\quad reproduced by us & 50.5$\pm$6.9 & 79.5$\pm$5.7\\
NeuroBridge~\cite{zhangNeuroBridgeBioInspiredSelfSupervised2026} & 63.2 & 89.9\\
\quad reproduced by us & 62.1$\pm$6.9 & 89.6$\pm$5.5\\
HVF~\cite{zhengLearningBrainRepresentation2026} & 75.7 & 94.6\\
LVBP~\cite{liuLeveragingVisualBlur2026} & 80.0 & 96.9\\
\quad reproduced by us & 80.0$\pm$5.0 & 96.7$\pm$1.7\\
\quad averaged after decoding & 70.9$\pm$5.7 & 92.8$\pm$2.7\\\midrule
Ours & 68.3$\pm$5.0 & 92.5$\pm$2.1\\
\quad averaged after decoding & \textbf{71.5}$\pm$4.7 & \textbf{93.7}$\pm$1.8\\\bottomrule
\end{tabular}
\caption{\textbf{Standard protocol.} Retrieval accuracy (\%) with the test EEG averaged over the 80 repetitions of each
image, as reported and as reproduced by us with the authors' code (one run each). Averaged after decoding, each
repetition is decoded on its own and the 80 embeddings are averaged (two runs for LVBP, five seeds for ours).}\label{tab:perf}
\end{minipage}\makeatother
\end{figure*}

We first decode every test trial on its own (Table~\ref{tab:single}). The published methods, run with the authors'
code, reach 5.0\%, 6.0\% and 14.1\% top-1. They were designed for and trained on averaged EEG. Our final model reaches
16.8\%, above LVBP, the strongest of them, in all ten subjects ($+2.7\pm1.2$ points), with fewer
parameters. To separate the architecture from the training recipe, we also train only the EEG networks of UBP,
NeuroBridge and LVBP in our pipeline, exactly as our model is trained (Table~\ref{tab:arch}). In each of the three training
regimes our model is the most accurate in all ten subjects, by $+3.9\pm0.7$, $+2.6\pm1.4$ and $+3.9\pm1.0$ points over
the stronger network, with less than half its parameters.

The standard benchmark averages the 80 repetitions of each test image before decoding (Table~\ref{tab:perf}). Here our
model reaches 68.3\%, above UBP and NeuroBridge, whose numbers we reproduce within 1.1 points. HVF and LVBP report
higher accuracies. Both align the EEG with features from several image encoders or several blurred versions of each
image, which is complementary to our approach and could be combined with it. LVBP's accuracy also depends on averaging the EEG before decoding. When each
repetition is decoded on its own and the 80 predictions are then averaged, LVBP reaches 70.9\% top-1 and our model
71.5\%. The same set of weights serves both protocols. On a single trial the dictionary projection keeps
only what the latency waveforms explain at each onset, and after averaging, when noise and neighbors are mostly gone, it
passes the clean response to image $i$ through. It needs a similar amount of computation per trial as UBP and NeuroBridge (Supp.~\ref{sec:supp_recipe}).

\begin{figure*}[t]\centering\includegraphics[width=\linewidth]{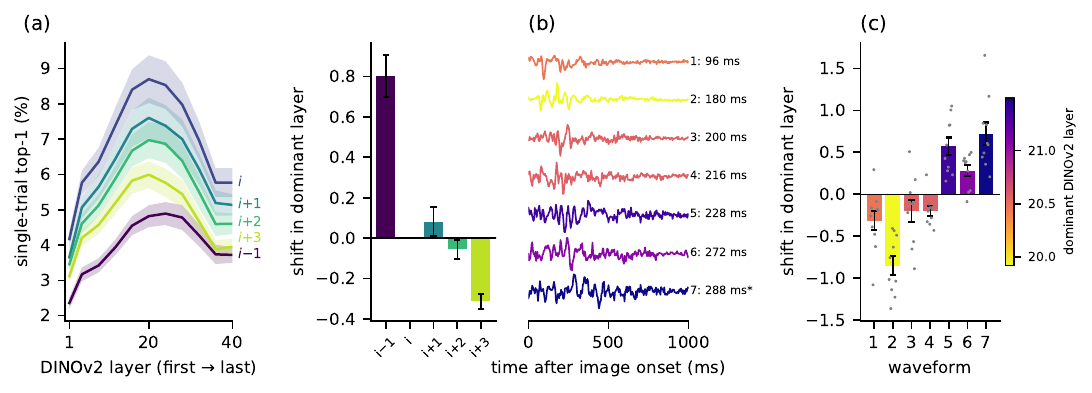}
\caption{\textbf{The model follows the visual hierarchy.} (a) Single-trial top-1 accuracy of a linear map from
the coefficients of the image at each position to the features of each DINOv2 layer (left), and the shift of each
position's dominant layer from that of image $i$ (right). (b) The seven latency waveforms, ordered by peak time and
colored by their dominant layer. (c) Shift of each waveform's dominant layer from the mean of the seven. Mean of ten
subjects $\pm$ s.e.m., gray dots single subjects. $^*$Peaks at 284--292\,ms in six subjects and 440\,ms in four.}
\label{fig:probe}\end{figure*}

To test a second dataset and a faster presentation rate, we train the same model on THINGS-EEG1~\cite{grootswagersHumanEEGRecordings2022}, where images appear every 100\,ms and each training image is shown once. We change only the spacing of the onsets in
$M$ and model nine images per window. We use the ten subjects with a complete test recording. Our model is again more
accurate than the EEG network of NeuroBridge, on single trials ($+2.1\pm0.6$ points) and on the average of the 12
test repetitions ($+4.2\pm3.1$ points), and on single trials it is the most accurate in all ten subjects
(Supp.~Table~\ref{tab:eeg1}).

\subsection{Interpreting the model}\label{sec:interp}
Every component of the model is a term of the signal representation, so each can be inspected directly (Supp.~\ref{sec:supp_dictionary}). A simple test shows that the projection separates the images. Each presentation of an image appears in five consecutive windows, each time at a different position and next to different images. If the projection isolates its response, the model should identify the image in the same way from all five windows. The embeddings of one presentation read from two windows are far more similar than those of two different images read from the same two windows (cosine 0.80 against 0.12, $p<10^{-14}$). Each test image, read only from the windows in which it appeared at one position, is reconstructed into nearly the same picture whether it appeared at the start of the window or 600\,ms later (Fig.~\ref{fig:posavg}).

Where an image falls in the window also decides which part of its response the window holds. Image $i{+}3$ appears 600\,ms into the window, so the window holds only the beginning of its response, while image $i{-}1$ appeared before the window started, so the window holds only its later part. Early visual areas, which encode simple features, respond before later areas, which encode the kind of object. A linear map from the image's coefficients to the features of each DINOv2 layer, fitted on training trials and scored on single test trials, shows this order (Fig.~\ref{fig:probe}a). We summarize each accuracy curve by its dominant layer, the mean of the layer numbers weighted by their accuracy above chance. At every position, decoding is best from the middle layers, but the previous image is decoded from later layers than image $i$ ($+0.8$ layers) and image $i{+}3$ from earlier ones ($-0.3$), in all ten subjects ($p<10^{-4}$, Supp.~\ref{sec:supp_domlayer}).

The latency waveforms show the same order inside the model. Nothing in training asks the waveforms to differ in content, and each one is defined only by its time course. Yet decoding from the coefficients of each waveform separately, waveforms that peak later have a later dominant layer (Fig.~\ref{fig:probe}b,c, rank correlation 0.76, $p<10^{-7}$, positive in all ten subjects). The model has divided the visual response into stages on its own. The waveforms peaking between 100 and 220\,ms carry relatively more of the low-level features of the first DINOv2 layers, and those peaking later more of the object-level features of the last layers. The waveforms are therefore an interpretable part of the model, a small set of time courses that each stand for a stage of visual processing. This matches the order in which deep network layers explain brain responses over time~\cite{cichyComparisonDeepNeural2016} and the parallel processing of successive images at different stages~\cite{kingHumanBrainEncodes2021}, here found from the model's own parameters in single trials.

\section{Discussion}\label{sec:discussion}
Decoders of visual EEG have so far read each window as the response to one image and treated everything else as noise.
We showed that much of this noise is made up of other images. A single window of fast presentation holds five images that can be decoded, the current one, the previous one and the next three, and a model built from how responses overlap reads all of them with
one readout.

This matters most where decoding is hardest. Real-time applications see each image once, and on single trials our model
is more accurate than every published model we evaluated, on two datasets recorded at different presentation rates. The standard benchmark averages 80 repetitions, which removes the neighbors with the noise and
hides much of what a window contains. We therefore recommend reporting single-trial results next to the averaged ones.

\paragraph{Limitations and outlook.} Each assumption is an approximation, and shared latency waveforms may be too simple
for data averaged over many repetitions. The same equations apply wherever events overlap at known times, and
constraining the sources by anatomy would also show where each image is processed.

\section{Conclusion}
Every EEG window recorded during fast presentation holds the responses to several images. We described this overlap with five assumptions and turned the resulting equations into a decoder. The model reads five images from a single window, where published decoders see only one. It is more accurate than every published model we evaluated on single trials, the setting that real-time decoding depends on, with fewer parameters. Because every component is a term of the equations, it also turns the decoder into a tool for studying how the brain keeps up with a rapid sequence of images. The brain may process several of these images without conscious access to them, which may be why the decoder can still distinguish images that the participant cannot. The brain processes several images at once, and so can a decoder.

{
    \small
    \bibliographystyle{ieeenat_fullname}
    \bibliography{main}
}

\clearpage
\setcounter{page}{1}
\maketitlesupplementary
\renewcommand{\dbltopfraction}{0.95}\renewcommand{\dblfloatpagefraction}{0.85}\setcounter{dbltopnumber}{3}
\renewcommand{\topfraction}{0.9}\renewcommand{\floatpagefraction}{0.85}

\section*{LLM Usage Statement}
We used Claude (Anthropic) to write and run code, to help prepare experiments, analyses and figures under the authors'
direction and review, to edit the text and related work, and to give feedback on the manuscript. It was not used for
the research ideas, hypotheses or conclusions. The authors take full responsibility for all content.

\FloatBarrier
\section{How many images does a window hold?}\label{sec:supp_leak}
One of our motivating questions arises from the idea that RSVP presents images to the subject faster than the brain can finish processing them. Consequently, multiple images should be decodable from the brain at once. The derivation of our architecture therefore began with a small pilot study aimed at (1) confirming this hypothesis and (2) quantifying it.

The number of decodable images naturally depends on window size. For comparability purposes we use the same settings as the state of the art. This means that for every trial $i$ we take the standard one-second window after its onset. On this window, we apply a cross-validated logistic regression to classify the image shown $k$ trials earlier or later (200 classes, chance
0.5\%, trained on three sessions and tested on the fourth, ten subjects).

\begin{table}[!htbp]\centering\small\setlength{\tabcolsep}{3pt}
\resizebox{\linewidth}{!}{%
\begin{tabular}{lccccccccccc}\toprule
Image & $i{-}5$ & $i{-}4$ & $i{-}3$ & $i{-}2$ & $i{-}1$ & $\mathbf{i}$ & $i{+}1$ & $i{+}2$ & $i{+}3$ & $i{+}4$ & $i{+}5$\\\midrule
$\times$ chance & 0.96 & 1.10 & 1.35 & 3.03 & 11.6 & \textbf{21.4} & 17.3 & 12.4 & 8.1 & 3.00 & 0.97\\
s.d. & 0.19 & 0.24 & 0.25 & 1.35 & 6.1 & 11.3 & 8.8 & 6.5 & 4.3 & 0.91 & 0.16\\\bottomrule
\end{tabular}}
\caption{\textbf{One window holds about five images.} Decodability of each image from trial $i$'s window ($\times$
chance), mean and s.d.\ over the ten subjects. The next image is almost as decodable as the current one. The spread is
between subjects, with every subject decoding all five images above chance (3.8--42$\times$).}\label{tab:leak}\end{table}

Table~\ref{tab:leak} shows that a window carries information about more images than the one it belongs to. The current image is decoded at $21.4\times$ chance, and the three images that follow it remain almost as strong ($17.3\times$, $12.4\times$, $8.1\times$), because they appear while the window is still running and their responses unfold within it. Looking back, the signal fades quickly. The previous image is still strong at $11.6\times$ chance, but two images back only the last part of the response is left ($3.0\times$), and further back it is close to chance. Image $i+4$ is just as weak ($3.0\times$), because it appears only 200 ms before the window ends, so the window catches just the start of its response. We conclude that five images, from $i - 1$ to $i+3$, are clearly decodable from a one-second window.

When only one of five overlapping responses is decoded, the other four act as noise. The usual method to deal with this noise is deconvolution. It estimates the average response to an image onset and subtracts this response for every neighbor in each window. We tested it on THINGS-EEG2 and concluded that this method does not remove the neighbors. In fact, not only are the neighbors just as decodable after deconvolution (paired t-tests, $p \geq 0.06$ at every lag), we also observe a drop in retrieval performance with ATM-S from 23.0\% to 21.1\% top-1 ($p = 0.022$). The reason is that deconvolution subtracts the same response for every image, while what identifies an image is how its response differs from the others. Subtracting an average leaves that difference in place, so the neighbors cannot be removed this way and have to be modeled instead. This further motivates our approach.

\FloatBarrier
\section{Training details}\label{sec:supp_aug}
\paragraph{Training on averages or single trials.} In THINGS-EEG2 every training image is shown four times to each subject. While training and testing on averaged repetitions in other models leads to a cleaner signal and improved performance, this is not exactly the case for us. Given our architecture, repetition averaging entails a trade-off.

On the one hand, the neighboring images become averaged with images of different identity thus diluting the positive effect that their identification could produce. On the other hand, averaging cleans up the signal of image $i$.

To study the effect of each of the settings we train models using each of these two paradigms (the first two rows of Table~\ref{tab:trainmix}).

Finally, to use this trade-off, we test one last setting, a mix of repetition averages and single trials. The aim is to combine the advantages of both settings, giving the model a cleaner signal of image $i$ while keeping the model's ability to learn the identities of the neighboring images. For this setting, in each batch, half of the rows are replaced by the mean of the four repetitions of the same image from the same subject, and the rest stay single trials. The model therefore sees two signal-to-noise regimes with the same targets. The neighbors' labels are used on the single-trial rows only, since each repetition of an image had different neighbors (final row of Table~\ref{tab:trainmix}).

Training on the mix gives the best accuracy across both protocols.

\begin{table}[!htbp]\centering\small\setlength{\tabcolsep}{3pt}
\resizebox{\linewidth}{!}{%
\begin{tabular}{lccccc}\toprule
 & \multicolumn{3}{c}{Single trial} & \multicolumn{2}{c}{80 repetitions}\\
Training data & Top-1 & Top-5 & $i{+}1$ & Top-1 & Top-5\\\midrule
Single trials only & 16.7$\pm$4.1 & 40.8$\pm$6.9 & 16.0$\pm$3.9 & 61.6$\pm$8.7 & 88.8$\pm$4.6\\
Averages only & 12.8$\pm$2.8 & 33.3$\pm$5.0 & 8.7$\pm$1.9 & 68.1$\pm$6.9 & 92.2$\pm$3.6\\
\textbf{Half and half (ours)} & \textbf{16.8}$\pm$3.9 & \textbf{41.0}$\pm$6.6 & \textbf{16.1}$\pm$3.9 & \textbf{68.3}$\pm$5.0 & \textbf{92.5}$\pm$2.1\\\bottomrule
\end{tabular}}
\caption{\textbf{Training data.} Our model trained on single trials only, on 4-repetition averages only, or on half of
each (the final model), with all other settings equal. Top-1/top-5 for image $i$, and single-trial top-1 for the next image
($i{+}1$). Mean $\pm$ s.d.\ over the ten subjects (three seeds each).}\label{tab:trainmix}\end{table}

\paragraph{Baselines.} The EEG networks of UBP, NeuroBridge and LVBP are trained, as in their papers, on the 4-repetition
averages of every training image, and additionally on single trials and on the same mix as our model. The augmentation
priors of the full NeuroBridge model are not part of its encoder and are not used. LVBP's network reads all 63 channels
and is trained with AdamW (learning rate $10^{-3}$, batch 1024).

\FloatBarrier
\section{Hyperparameter Settings}\label{sec:supp_settings}
For hyperparameter tuning we held out 10\% of the training concepts as a validation set. These concepts are never used in training, so validation is zero-shot, like the test.

\paragraph{Bayesian search.} The per-subject model's settings were chosen by a Bayesian search (Optuna TPE, 50 trials,
Hyperband pruning between 5 and 40 epochs) on subjects 1, 5 and 9, scored by validation top-1. The search space was as follows.
\begin{itemize}
    \item sources $R$ 4--32,
    \item filters per source $F$ 2--8,
    \item filter length 13/25/51 samples,
    \item latency waveforms $Q$ 4--16,
    \item noise waveforms $L$ 0--16,
    \item images modeled 5 or 7,
    \item the regularization constant $\lambda$ (Eq.~\eqref{eq:solve}) $10^{-4}$--$10^{-1}$,
    \item feature dropout 0--0.5,
    \item learning rate of the network $3\times10^{-4}$--$3\times10^{-3}$,
    \item learning rate of the image coefficients $3\times10^{-3}$--$3\times10^{-2}$, which were learned parameters in the version searched and are computed by the dictionary projection in the final model,
    \item train-mix ratio (averages vs single trials) (0.5--1) (see experiment on Table~\ref{tab:trainmix}).
\end{itemize}
 About fifteen settings scored within one point of the best (19.0--19.9 validation top-1, 1,654-way), showing a broad plateau. A functional ANOVA on the 50 trials attributes most of the variance to the train-mix ratio (31\%, best 0.5--0.6), the number of noise waveforms (14\%, best 11--16) and the learning
rates (25\%), and $R$ mattered least (2\%). The best three settings were re-checked with two seeds on subjects 2, 6 and 10,
not used in the search, and the winner kept, with $Q{=}7$, $L{=}12$, 13-sample filters, $\lambda=6.6\times10^{-4}$, dropout
0.44, averaged fraction 0.5.
\paragraph{Front end for pooled training.} The search ran per subject, where a small front end ($R{\times}F=13{\times}6$)
was best. Pooling ten subjects gives ten times the data, so we re-checked capacity on the pooled model: $13{\times}6$,
$24{\times}8$ and $32{\times}12$, scored on a high-SNR validation that averages each held-out image's 40 recordings (ten
subjects $\times$ four) in the shared source space, since four-repetition validation could not rank models for the
80-repetition test. $32{\times}12$ was best and kept, and a larger front end ($48{\times}16$) and an isolated-window readout
did not improve the fine-tuned model on validation.
\paragraph{Training.} For the targets we compared DINOv2-giant hidden states, averaged over tokens, at layers 10, 16, 20, 24, 27 and 34 of 40
(validation top-1 29.6, 34.8, 34.7, 32.2, 29.5, 23.4, against 19.9 for CLIP ResNet-50). Layers 16 and 20 tie on validation, and we kept 20.
Pooled training runs for 12 epochs. In fine-tuning, validation peaks after 2--6 epochs and then declines, so
we use 4 epochs. Four other fine-tuning variants (learning rate $10^{-4}$ for 25 epochs, with or without averages only,
$3\times10^{-4}$ for 12 epochs, and training only the unmixing and readout) tie with it on validation (17.7--18.1 top-1 for
all five), so the simplest was kept.
All our models are trained with Adam (learning rate $1.45\times10^{-3}$, batch 64) and a contrastive loss with a learned temperature, initialized at 0.07.

\paragraph{Baselines.} The EEG networks of UBP, NeuroBridge and LVBP use their own losses and optimizers. The epochs of
UBP and NeuroBridge were chosen on their own validation curves (UBP 36 pooled + 25 fine-tuning on averages and 9 + 6 on
single trials, NeuroBridge 31 + 25 and 8 + 6). LVBP's network is trained per subject without pooling, for 200 epochs on
averages, as by its authors, and 13 on single trials. The mix uses the single-trial epochs.

\FloatBarrier
\section{Ablations}\label{sec:supp_ablations}
\begin{table*}[!tp]\centering\small\setlength{\tabcolsep}{3pt}
\begin{tabular}{clccc}\toprule
 & Change & Single trial & 80 rep. & Model\\\midrule
A1 & Final model & 16.8$\pm$3.9 & 68.3$\pm$5.0 & final\\
A2 & No dictionary projection (readout on $h$) & 9.6$\pm$2.4 & 51.8$\pm$6.9 & final, label $i$\\
A3 & Learned matrix in place of $A$ & 16.8$\pm$4.0 & 69.5$\pm$4.9 & final\\
A4 & No temporal filters ($R{=}32$, $F{=}1$) & 12.7$\pm$2.7 & 50.1$\pm$6.6 & final\\
A5 & No filters, same channel count ($R{=}384$, $F{=}1$) & 11.0$\pm$3.0 & 49.1$\pm$7.6 & final\\\midrule
B1 & Images modeled 1 / 2 / 3 / 5 & 11.1 / 11.1 / 11.1 / 11.4 & 60.5 / 59.7 / 59.0 / 61.0 & per subject$^\dagger$\\
B2 & Neighbors with shared response only & 11.4 & 60.0 & per subject$^\dagger$\\
B3 & No noise term ($L{=}0$), 5 images & 11.4 & 61.2 & per subject$^\dagger$\\
B4 & Own waveforms per source instead of shared & \multicolumn{2}{c}{$-2.3$ val.\ top-1} & per subject$^\ddagger$\\
B5 & Images modeled 5 vs 7 & \multicolumn{2}{c}{$-0.7$ val.\ top-1 (tie in top-5)} & per subject$^\ddagger$\\\bottomrule
\end{tabular}
\caption{\textbf{Ablations}, top-1 (\%), image $i$. Rows A, the final model of the main paper, mean $\pm$ s.d.\ over the ten subjects, three seeds each (``label $i$'' is its
variant trained with image $i$'s label only). Rows B, ablations of the structure of the equation, run in the
per-subject setting of the hyperparameter search (Supp.~\ref{sec:supp_settings}), where many variants can be compared at low cost.
$^\dagger$Trained per subject without pooling ($13{\times}6$ front end, 25 epochs, subjects 1, 5 and 9, one seed). Each
row is compared with the same setting with five images modeled (11.4 and 61.0). $^\ddagger$Validation top-1 within
the search, two seeds. The absolute values of rows B are lower than the final model's because the setting is smaller,
and only differences within it are compared.}\label{tab:ablations}\end{table*}
Removing the temporal filters, so that each source gives a single channel, costs 4.1 points on single trials (A1 to A4, 16.8\% to
12.7\%) and 18.2 points with 80 repetitions (68.3\% to 50.1\%). Giving the model as many channels as the final model
but no filters (384 sources, A5) does not recover it (11.0\% and 49.1\%), so what matters is the filtering, not the number
of channels. Removing the dictionary projection (A2) is the largest single loss on single trials. Image $i$ falls from 16.2\% to 9.6\% on single trials and from 67.4\%
to 51.8\% with 80 repetitions (both models trained with image $i$'s label only), and the neighbors fall to chance. Sharing the latency waveforms across sources is better than
giving each source its own (B4), which multiplies the unknowns of every projection. For retrieval of image $i$ alone, the number of
images modeled makes little difference (11.1--11.4\% in the per-subject setting, B1 and B5), because the projection's gain on image $i$ comes from separating it from the
noise as much as from its neighbors. A learned matrix in place of $A$ (A3), initialized from it and trained with the labels of all five images, matches the final model (16.8\% and 69.5\%, next image 16.3\% against 16.1\%). Without the neighbors' labels, only the projection itself separates the images (Sec.~\ref{sec:multi}). Modeling the neighbors is what makes them decodable (Sec.~\ref{sec:multi}), and learning their labels does not
cost the current image under the standard protocol. The final model reaches 68.3$\pm$5.0\% at 80 repetitions with the
labels of all five images and 67.4$\pm$6.6\% with image $i$'s label only (paired difference $+0.9\pm3.0$, not
significant, $p=0.36$). Combining an image's coefficients from the five windows that contain one presentation does not
improve its retrieval either (16.2\% against 16.9\% from its own window, on the presentations whose five windows all exist, lower in every subject), since the windows give
largely the same view of one response, consistent with the agreement of Sec.~\ref{sec:interp}.

Together, the ablations show that the model's accuracy rests on two components. The temporal filters give each source
several channels to separate, and the dictionary projection uses these channels to assign every part of the signal to
one image. Replacing the projection with a learned matrix gives the same accuracy, so its benefit comes from the
structure it imposes, not from avoiding training.

\FloatBarrier
\section{The learned dictionary}\label{sec:supp_dictionary}
\begin{figure*}[!tp]\centering\includegraphics[width=\linewidth]{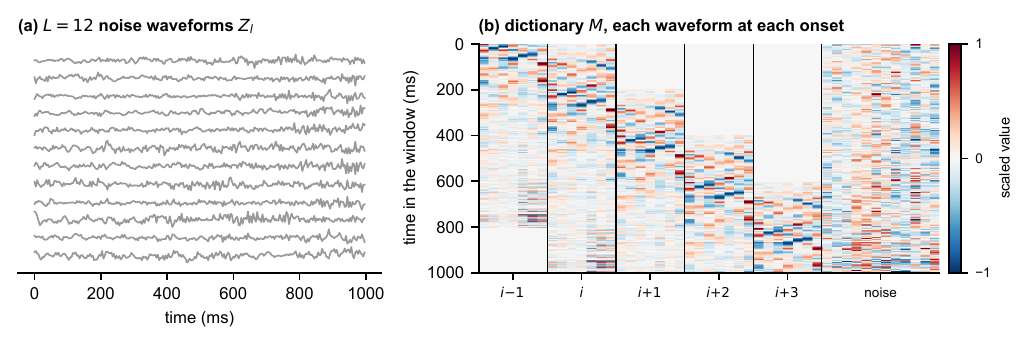}
\caption{\textbf{The dictionary $M$ of the final model} (subject 8). (a) The twelve noise waveforms $Z_l$, each scaled to its maximum. (b) $M$ as a matrix: each of the seven latency
waveforms $P_q$ (Fig.~\ref{fig:probe}b) placed at the onset of each of the five images (cut to the window, with gray marking
the time before the onset or after the window),
followed by the noise waveforms. Colors show each column scaled to its largest absolute value, red positive and blue negative.}\label{fig:dictionary}\end{figure*}
The waveforms are learned through the dictionary projection (Fig.~\ref{fig:dictionary}) by the contrastive loss alone, with no constraint on smoothness. They are not
event-related potentials but a basis in which the images' responses in the feature channels can be separated.

\FloatBarrier
\section{Dominant DINOv2 layer}\label{sec:supp_domlayer}
For every accuracy curve over DINOv2 layers in Fig.~\ref{fig:probe}, the dominant layer is the mean of the layer
numbers (1, 4, 8, \dots, 40) weighted by the accuracy above chance at each layer. It summarizes the curve without
choosing which layers count as early or late. Tables~\ref{tab:domlayer_pos} and~\ref{tab:domlayer_wave} give the values
behind Fig.~\ref{fig:probe}.

\begin{table}[!htbp]\centering\small\setlength{\tabcolsep}{4pt}
\begin{tabular}{lccc}\toprule
Position & Dominant layer & Shift from image $i$ & Subjects\\\midrule
$i{-}1$ & 21.45$\pm$0.75 & $+$0.80$\pm$0.32 & 10 later\\
$i$     & 20.65$\pm$0.58 & -- & --\\
$i{+}1$ & 20.73$\pm$0.57 & $+$0.08$\pm$0.23 & 5 later\\
$i{+}2$ & 20.59$\pm$0.58 & $-$0.06$\pm$0.15 & 5 earlier\\
$i{+}3$ & 20.33$\pm$0.55 & $-$0.31$\pm$0.11 & 10 earlier\\\bottomrule
\end{tabular}
\caption{\textbf{Dominant DINOv2 layer per position} (Fig.~\ref{fig:probe}a). Mean $\pm$ s.d.\ over ten subjects,
each the mean of two seeds. Shifts at $i{-}1$ and $i{+}3$, $p<10^{-4}$ (paired $t$-test).}\label{tab:domlayer_pos}
\end{table}

\begin{table}[!htbp]\centering\small\setlength{\tabcolsep}{4pt}
\resizebox{\linewidth}{!}{%
\begin{tabular}{ccccc}\toprule
Waveform & Peak (ms) & Dom.\ layer & Shift & Top-1 (\%)\\\midrule
1 & 96  & 20.45$\pm$0.55 & $-$0.32$\pm$0.37 & 3.4$\pm$0.8\\
2 & 180 & 19.92$\pm$0.64 & $-$0.85$\pm$0.35 & 2.8$\pm$0.5\\
3 & 200 & 20.57$\pm$0.83 & $-$0.20$\pm$0.41 & 3.9$\pm$1.0\\
4 & 216 & 20.57$\pm$0.59 & $-$0.20$\pm$0.18 & 3.7$\pm$0.9\\
5 & 228 & 21.34$\pm$0.63 & $+$0.57$\pm$0.33 & 3.8$\pm$0.8\\
6 & 272 & 21.05$\pm$0.69 & $+$0.28$\pm$0.21 & 3.6$\pm$0.9\\
7 & 288$^*$ & 21.49$\pm$0.82 & $+$0.72$\pm$0.43 & 3.2$\pm$0.8\\\bottomrule
\end{tabular}}
\caption{\textbf{Dominant DINOv2 layer per latency waveform} (Fig.~\ref{fig:probe}b,c), from the coefficients of image
$i$ on each waveform alone. Dom.\ layer, dominant layer. Shift, from the mean of the seven waveforms of the same subject. Top-1, accuracy at the best layer. Mean $\pm$ s.d.\ over ten subjects, one seed. $^*$Median. It peaks at 284--292\,ms in six
subjects and 440\,ms in four. Rank correlation of peak time and dominant layer $0.76\pm0.14$, positive in all ten
subjects ($p<10^{-7}$).}\label{tab:domlayer_wave}\end{table}

\FloatBarrier
\section{THINGS-EEG1}\label{sec:supp_eeg1}
THINGS-EEG1~\cite{grootswagersHumanEEGRecordings2022} shows images every 100\,ms, twice as fast as THINGS-EEG2, and each
training image only once, so no repetition averages are available for training. We train the same model with only the
spacing of the onsets in $M$ changed, and model nine images per window. We use the ten subjects with a complete test
recording, in which each test image is shown 12 times. The EEG networks of UBP, NeuroBridge and LVBP are trained on
single trials in our pipeline. Table~\ref{tab:eeg1} shows that our model is the most accurate on single trials and on
the average of the 12 repetitions.

\begin{table}[!htbp]\centering\small\setlength{\tabcolsep}{3pt}
\resizebox{\linewidth}{!}{%
\begin{tabular}{lcccc}\toprule
& \multicolumn{2}{c}{Single trials} & \multicolumn{2}{c}{12 repetitions}\\
EEG network & Top-1 & Top-5 & Top-1 & Top-5\\\midrule
UBP & 3.5$\pm$0.9 & 13.0$\pm$2.4 & 14.9$\pm$3.0 & 39.8$\pm$5.9\\
NeuroBridge & 3.5$\pm$0.9 & 12.8$\pm$2.3 & 15.2$\pm$3.7 & 40.3$\pm$6.0\\
LVBP & 2.3$\pm$1.4 & 9.0$\pm$4.5 & 6.7$\pm$5.2 & 21.1$\pm$13.8\\\midrule
Ours & \textbf{5.6}$\pm$1.1 & \textbf{18.7}$\pm$3.0 & \textbf{19.3}$\pm$2.5 & \textbf{48.7}$\pm$4.5\\\bottomrule
\end{tabular}}
\caption{\textbf{THINGS-EEG1} (100\,ms per image). Accuracy (\%) for image $i$, ten subjects, three seeds, networks
trained on single trials in our pipeline. Across seeds, no number of ours varies by more than 2.0 points. Ours is
the most accurate in all ten subjects on single trials ($p<10^{-5}$ against each network) and on the 12 repetitions
against UBP and LVBP ($p<10^{-4}$), and in eight of ten against NeuroBridge ($p<0.01$).}\label{tab:eeg1}\end{table}

\FloatBarrier
\section{Published EEG networks in our pipeline}\label{sec:supp_recipe}
\begin{table*}[!t]\centering\small\setlength{\tabcolsep}{6pt}
\begin{tabular}{llcccc}\toprule
 & & \multicolumn{2}{c}{Single trial} & \multicolumn{2}{c}{80 repetitions}\\
Model & Trained on & Top-1 & Top-5 & Top-1 & Top-5\\\midrule
UBP network & averages & 7.6$\pm$2.2 & 22.7$\pm$4.9 & 68.6$\pm$5.3 & 92.4$\pm$3.9\\
NeuroBridge network & averages & 8.8$\pm$2.6 & 25.7$\pm$5.7 & 74.0$\pm$3.6 & 94.6$\pm$2.1\\
LVBP network & averages & 5.3$\pm$1.3 & 17.0$\pm$3.4 & 52.1$\pm$5.6 & 81.1$\pm$3.9\\
Ours & averages & 12.8$\pm$2.8 & 33.3$\pm$5.0 & 68.1$\pm$6.9 & 92.2$\pm$3.6\\\midrule
UBP network & single trials & 11.8$\pm$3.4 & 32.2$\pm$6.6 & 52.6$\pm$5.7 & 83.8$\pm$4.9\\
NeuroBridge network & single trials & 15.3$\pm$4.8 & 38.6$\pm$8.7 & 66.3$\pm$5.6 & 91.1$\pm$3.4\\
LVBP network & single trials & 13.0$\pm$3.1 & 34.5$\pm$6.3 & 55.7$\pm$5.8 & 86.8$\pm$3.0\\
Ours & single trials & 17.9$\pm$4.1 & 43.4$\pm$6.8 & 43.1$\pm$11.5 & 74.9$\pm$10.5\\\midrule
UBP network & mix & 10.1$\pm$3.1 & 28.4$\pm$6.3 & 62.6$\pm$5.9 & 89.7$\pm$4.1\\
NeuroBridge network & mix & 12.3$\pm$4.0 & 33.0$\pm$7.8 & 68.4$\pm$4.6 & 92.6$\pm$2.8\\
LVBP network & mix & 10.6$\pm$2.7 & 29.3$\pm$5.6 & 61.2$\pm$4.9 & 89.7$\pm$2.4\\
Ours & mix & 16.2$\pm$3.9 & 40.1$\pm$6.7 & 67.4$\pm$6.6 & 91.9$\pm$3.2\\\bottomrule
\end{tabular}
\caption{\textbf{Published EEG networks in our pipeline.} Accuracy (\%) of the EEG networks of UBP, NeuroBridge and LVBP and
of our model, all trained in our pipeline with the label of the current image only, on 4-repetition averages, single
trials or a mix with half of each batch averaged. Ours differs from Table~\ref{tab:trainmix}, where it is trained with the labels of all five images, except on averages, where neighbor labels cannot be used, so that row is the same model.}\label{tab:recipe}\end{table*}
\begin{table*}[!tp]\centering\small\setlength{\tabcolsep}{2.5pt}
\begin{tabular}{lccccccc}\toprule
Method & PixCorr & SSIM & Alex(2) & Alex(5) & Incep. & CLIP & SwAV$\downarrow$\\\midrule
ATM-S~\cite{liVisualDecodingReconstruction2024} (reported, subject 8) & 0.160 & \textbf{0.345} & 0.776 & 0.866 & 0.734 & \textbf{0.786} & 0.582\\
Ours, subject 8 & \textbf{0.180} & 0.340 & \textbf{0.802} & \textbf{0.885} & \textbf{0.756} & \textbf{0.786} & \textbf{0.532}\\\midrule
Ours, 10 subjects & 0.145$\pm$.016 & 0.329$\pm$.008 & 0.780$\pm$.016 & 0.861$\pm$.016 & 0.737$\pm$.025 & 0.779$\pm$.015 & 0.543$\pm$.014\\\bottomrule
\end{tabular}
\caption{Image reconstruction, 80 repetitions (SwAV is a distance, lower is better). Ours uses one seed, and the ten-subject row is mean
$\pm$ s.d.\ over subjects. ATM reports no spread.}\label{tab:recon}\end{table*}
UBP~\cite{wuBridgingVisionBrainGap2025} and NeuroBridge~\cite{zhangNeuroBridgeBioInspiredSelfSupervised2026} are complete
methods. Each consists of an EEG network, which maps a window to an embedding, and of several additions around it. UBP
blurs the images according to an uncertainty prior before computing their embeddings, and NeuroBridge augments the
images and the EEG with priors inspired by human perception and aligns both through learned projectors. Each method also
has its own preprocessing, image targets, training schedule and epoch selection.

These additions make it hard to tell whether a difference in accuracy comes from the architecture or from the rest of
the method. As an ablation, we therefore take only the EEG network of each method and train it in our pipeline, under
exactly the same conditions as our model. The network is the one released by the authors, a linear layer followed by a
residual block and layer normalization, applied to the 17 posterior channels both papers use. LVBP's network reads all 63 channels. NeuroBridge's version adds
linear projectors on both sides. The networks keep their own loss and optimizer, and everything else is ours, namely
the preprocessed EEG, the DINOv2 targets, the validation split used to choose the number of epochs, and the pre-training
on all subjects followed by fine-tuning on each subject. They are trained on 4-repetition averages, as published, on
single trials, and on the same mix of the two as our model.

This ablation shows that our model is more accurate than the published methods and also than their EEG networks alone,
in every training regime on single trials (Table~\ref{tab:arch}). It also produces an observation that we leave for
future work. In our pipeline the EEG networks are more accurate than the complete methods, on single trials (8.8\%
against 6.0\% for NeuroBridge, both trained on averages) and on the standard benchmark (74.0\% against 63.2\%). The
authors' own ablations report the networks without their additions at 42.1\% (UBP) and 40.5\% (NeuroBridge). The
additions therefore help in the authors' pipeline, while our pipeline helps the networks more. The DINOv2 targets and
the pre-training on all subjects are likely contributors, but we have not separated their effects.

On single trials our model is the most accurate in every regime. On the standard benchmark, NeuroBridge's network trained
on averages is the most accurate, consistent with Sec.~\ref{sec:discussion}, where our model's advantage is largest on single trials.

One NeuroBridge network per image position (48.4\,M parameters in total) reaches
5.9 / 15.3 / 14.2 / 12.5 / 10.1\% top-1 at $i{-}1$ to $i{+}3$, still below our single model (Fig.~\ref{fig:neighbours}).

At inference our model has 4.1\,M parameters (12.1\,k for the equation, the rest for the
readout), against 9.7\,M for NeuroBridge's encoder and 8.9\,M for UBP's, with similar compute per single trial (20.8,
19.3 and 17.8\,MFLOPs). Our single model decodes all five images.

\FloatBarrier
\section{Image reconstruction}\label{sec:supp_recon}
\begin{figure*}[!t]\centering\includegraphics[width=0.6\linewidth]{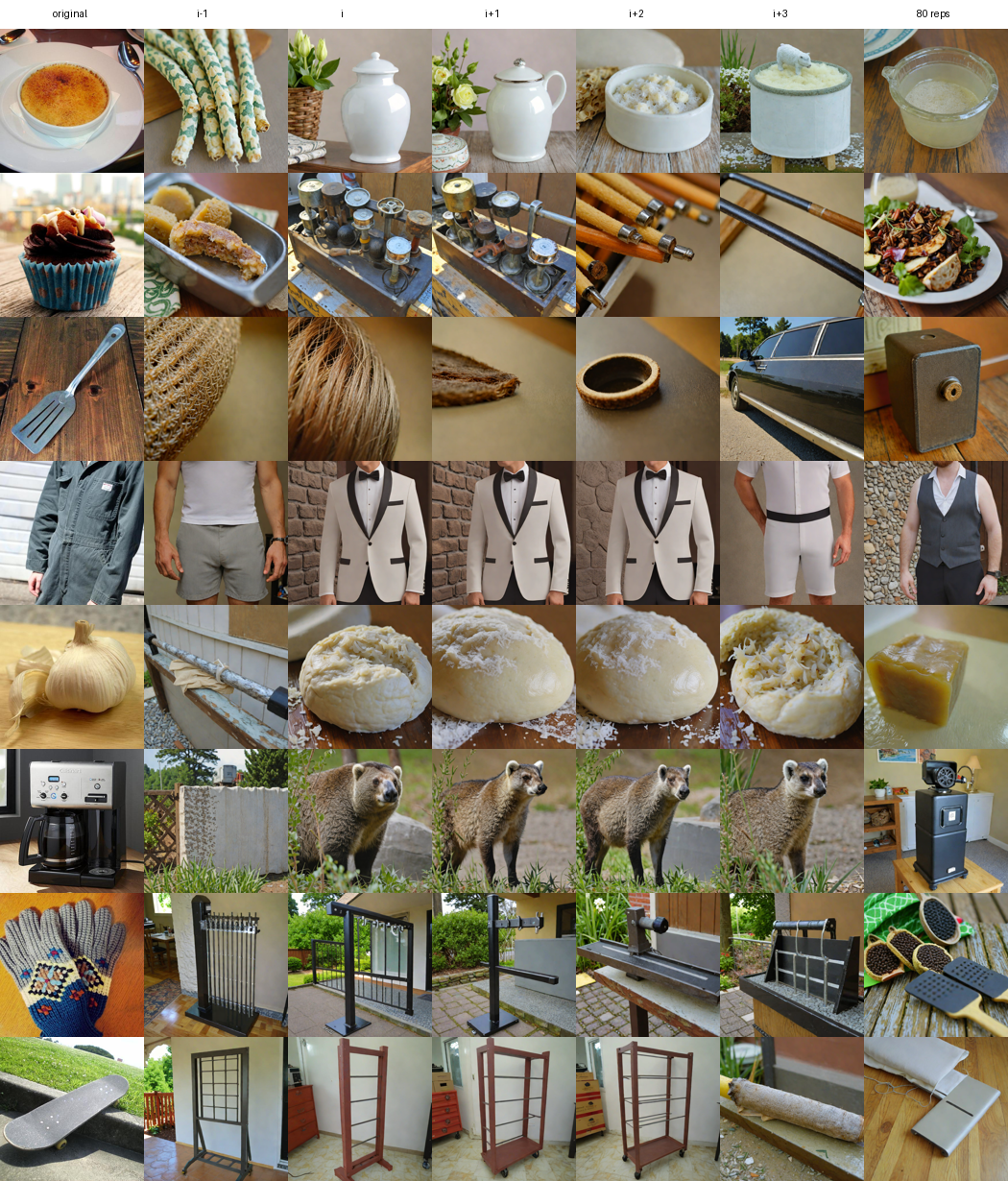}
\caption{\textbf{The same image presentation over time.} Each row is one presentation of a test image (left),
reconstructed from each of the five consecutive windows that contain it, at positions $i{-}1$ to $i{+}3$ (single
trials), and from the 80-repetition average for reference (right). The eight pairs of subject and image are chosen at
random. Reconstructions use SDXL-Turbo and IP-Adapter, as in ATM~\cite{liVisualDecodingReconstruction2024}.}\label{fig:interp}\end{figure*}

As in ATM~\cite{liVisualDecodingReconstruction2024}, reconstructions use SDXL-Turbo with an IP-Adapter conditioned on a CLIP
ViT-H image embedding (4 steps, no guidance, 512\,px). Our model predicts DINOv2 embeddings, and contrastive predictions
are not on the distribution of real image embeddings, so the generator input is the mean of the training images' CLIP
embeddings, weighted by $\mathrm{softmax}(\tau\cdot\text{similarity})$ to the prediction in the model's own space, with
$\tau$ chosen on held-out training images. Test images are never among them.

Table~\ref{tab:recon} uses the metrics and protocol of ATM, with the test EEG averaged
over 80 repetitions. Our model matches or exceeds ATM-S on six of the seven metrics for the subject ATM reports.

Each presentation of an image appears in five consecutive windows,
at positions $i{-}1$ to $i{+}3$. Fig.~\ref{fig:interp} reconstructs one presentation from each of these windows, for
eight random pairs of subject and image. The five windows give consistent reconstructions of one presentation, although
on single trials these often show a related or wrong object.

To separate time from signal strength (Sec.~\ref{sec:interp}), we
regressed each reconstruction's structural similarity (SSIM above a shuffled floor) on its semantic similarity (two-way
CLIP) and its position, per subject. At matched semantic similarity, late positions have no more structure than
position $i$ (coefficient $-0.0004$, $p=0.89$, PixCorr $-0.0048$, $p=0.19$, paired over ten subjects).

\end{document}